\documentclass[%
 superscriptaddress,
 twocolumn,
 nofootinbib,
 amsmath,amssymb,
 aps,
 prd,
]{revtex4-1}

\usepackage[colorlinks]{hyperref}
\usepackage{tabularx}
\usepackage{graphicx}
\usepackage{amssymb,amsmath,bm,tensor,braket}
\usepackage[dvipsnames]{xcolor}
\usepackage[varg]{txfonts}
\usepackage{enumerate}
\usepackage{mathtools}
\usepackage{tensor}
\usepackage[capitalize]{cleveref}
\usepackage[utf8]{inputenc}
\usepackage[normalem]{ulem}
\usepackage{dcolumn}
\usepackage{soul}
\usepackage{subfigure}
\usepackage[normalem]{ulem}

\usepackage{comment}
\usepackage{siunitx}

\newcommand{\dd}{\mathrm{d}}
\newcommand{\ii}{\mathrm{i}}
\newcommand{\ee}{\mathrm{e}}
\newcommand{\p}{^{\prime}}

\begin{document}

\preprint{APS/123-QED}

\title{Superradiant instabilities of massive bosons around exotic compact objects}

\author{Lihang Zhou}
\affiliation{Department of Astronomy, School of Physics, Peking University, Beijing 100871, China}%
\affiliation{Kavli Institute for Astronomy and Astrophysics, Peking University, Beijing 100871, China}

\author{Richard Brito}%
\affiliation{CENTRA, Departamento de F\'isica, Instituto Superior T\'ecnico -- IST, Universidade de Lisboa -- UL, Avenida Rovisco Pais 1, 1049 Lisboa, Portugal}

\author{Zhan-Feng Mai}\email[Corresponding author: ]{zhanfeng.mai@gmail.com}
\affiliation{Kavli Institute for Astronomy and Astrophysics, Peking University, Beijing 100871, China}%

\author{Lijing Shao}\email[Corresponding author: ]{lshao@pku.edu.cn}
\affiliation{Kavli Institute for Astronomy and Astrophysics, Peking University, Beijing 100871, China}%
\affiliation{National Astronomical Observatories, Chinese Academy of Sciences, Beijing 100012, China}

\date{\today}

\begin{abstract}
Superradiantly unstable ultralight particles around a classical rotating black
hole (BH) can form an exponentially growing bosonic cloud, which have been shown to provide an
astrophysical probe to detect ultralight particles and constrain their
mass. However, the classical BH picture has been questioned, and
different theoretical alternatives have been proposed. Exotic compact objects (ECOs)
are horizonless alternatives to BHs featuring a reflective surface (with a
reflectivity $\mathcal{K}$) in place of the event horizon. In this work, we
study superradiant instabilities around ECOs, particularly focusing on the
influence of the boundary reflection. We calculate the growth rate of
superradiant instabilities around ECOs, and show that the result can be
related to the BH case by a correction factor $g_{\mathcal{K}}$, for which we find an
explicit analytical expression and a clear physical interpretation.
Additionally, we consider the time evolution of superradiant instabilities and
find that the boundary reflection can either shorten or prolong the growth
timescale. As a result, the boundary reflection alters the superradiance
exclusion region on the Regge plane, potentially affecting constraints on the mass of
ultralight particles. For a mildly reflective surface
($|\mathcal{K}|\lesssim 0.5$), the exclusion region is not substantially
changed, while significant effects from the boundary reflection can occur for an
extreme reflectivity ($|\mathcal{K}|\gtrsim0.9$).
\end{abstract}

\maketitle

\allowdisplaybreaks

\section{Introduction}

Ultralight bosons have been proposed by different theories as elementary
particles beyond the Standard Model of particle physics. Examples include (i)
the QCD axion, introduced to solve the strong charge-parity problem
\cite{PhysRevLett.38.1440,Weinberg:1977ma,Wilczek:1977pj}, (ii) a plenitude of axion-like particles (ALPs),
predicted by the string theory and called an ``axiverse''
\cite{PhysRevD.81.123530,arvanitaki2011exploring}, and (iii) dark photons
\cite{Goodsell_2009}. These ultralight bosons, which would naturally couple weakly to baryonic matter, have been shown to be promising dark matter candidates
\cite{PhysRevD.89.083536, PhysRevD.95.043541,Antypas:2022asj}.

In addition to some ground-based experiments (see e.g.
Refs.~\cite{Wang:2021dfj,Antypas:2022asj}), astrophysical environments, such as the vicinity of black holes (BHs), also provide natural testbeds for detecting ultralight particles. This relies on a mechanism called BH superradiance (for a comprehensive review, see Ref.~\cite{Brito:2015oca}). Consider a field of ultralight particles with mass $\mu$, located near a rotating BH. When the
Compton wavelength of the particle is comparable to the horizon radius of the
BH, the particle field can form quasi-bound states around the BH and
extract energy and angular momentum effectively from
the BH if the following superradiance condition is satisfied
\begin{equation}
\omega_R<\frac{ma}{2Mr_+},
\label{eq:SR condition}
\end{equation}
where $\omega_R$ is the real part of the frequency of the massive field, which
is typically close to the mass $\mu$ of the ultralight particle; $m$ is the magnetic quantum number; and $a,\ M,\ r_+$ are the spin, mass and horizon
radius of the BH, respectively. An intuitive understanding of this condition is that superradiance occurs whenever the angular velocity of the field, $\omega_R/m$, is less than that of the spacetime $a/2Mr_+$. When the superradiance condition is met, the bosonic field can turn unstable as more and more particles are produced from the extracted energy and angular momentum. These particles remain bounded to the BH by gravity and form an exponentially growing bosonic cloud. This phenomenon is the so-called superradiant instability.
Superradiant instabilities of BHs have been thoroughly studied using
perturbation theory but also numerical relativity, with studies including the computation of the unstable eigenfrequencies~\cite{zouros1979, detweiler1980klein,
furuhashi2004instability, Cardoso:2004nk, Cardoso:2005vk,dolan2007instability, pani2012perturbations,Brito:2013wya,Dolan:2018dqv,
baumann2019spectra, bao2022improved, Dias:2023ynv}, linear and non-linear time evolutions of the instability~\cite{PhysRevD.87.124026,Witek2013,brito2015black,East2018}, the understanding of nonlinear effects such as the ``bosenova'' or scalar emission induced by self-interactions~\cite{Yoshino:2012kn,
Yoshino_2015,Baryakhtar:2020gao,Omiya:2022gwu}, and the computation of gravitational wave (GW) emission by the bosonic cloud~\cite{Yoshino:2013ofa,
arvanitaki2015discovering, brito2015black,PhysRevD.96.035019,Brito:2017wnc,Brito:2017zvb, Siemonsen:2019ebd, siemonsen2023superrad}.  In particular, it has been shown that superradiant instabilities can spin down rotating BHs and leave exclusion regions on the
BH spin-mass plane (Regge plane). Since these regions are related to the mass of ultralight
particles, they can be used to constrain the latter through the measurement of
the spin and mass of astrophysical BHs~\cite{arvanitaki2011exploring,
arvanitaki2015discovering,brito2015black, cardoso2018constraining, arvanitaki2017black}. 

In this paper, we shall not restrict our discussion to BHs. Classical BHs,
namely, Kerr BHs, as a solution of Einstein's general relativity (GR), have
singularities at $r=0$, which are hidden within event horizons. However, this classical
BH picture leads to puzzles \cite{Cardoso:2019rvt}, for example, the
information paradox of evaporating BHs \cite{PhysRevD.14.2460,Marolf:2017jkr}.
In different contexts, including some quantum gravity candidates, exotic compact
objects (ECOs) as horizonless alternatives to classical BHs, have been proposed
and studied. Examples include fuzzballs in string theory
\cite{Mathur:2005zp, SKENDERIS2008117}, boson stars and oscillatons
\cite{Visinelli:2021uve}, gravitational condensate stars, i.e. gravastars
\cite{Mazur:2001fv} and wormholes \cite{Visser:1995cc, PhysRevD.76.024016}. A more comprehensive list of proposed ECOs can be found in
Refs.~\cite{Cardoso:2017njb,Cardoso:2019rvt}.

ECOs have also been dubbed as ``BH mimickers'' \cite{cardoso2016gravitational}.  They have
no event horizon, and their surfaces are reflective, in contrast with the BH
event horizon which only allows particles and waves to fall inwards. In a simple
phenomenological model, widely used in the literature, one can assume the presence of
a boundary at $r_0\gtrsim r_+$, with a reflectivity $\mathcal{K}$, while the
external spacetime is still described by the usual Kerr geometry. The modified
inner boundary condition may result in colorful phenomenology, including
modified quasinormal modes (QNMs) \cite{PhysRevD.77.124044, macedo2018spectral,
maggio2017exotic, maggio2019ergoregion}, ergoregion instabilities
\cite{1978CMaPh..62..247F, Moschidis:2016zjy, PhysRevD.77.124044,
maggio2017exotic, maggio2019ergoregion} and GW ``echo'' signals \cite{Mark:2017dnq,
Cardoso:2017njb, Cardoso:2019rvt, Burgess:2018pmm}, to name a few examples. Although there
is no definite proof of the existence of BH horizons \cite{PhysRevD.77.124044}, the advent of GW detection and the improvements
in its precision may provide a unique opportunity to probe physics at the
near-horizon scale and constrain to very high precision the existence of a reflecting surface (see Refs.~\cite{Cardoso:2019rvt,Maggio:2021ans} for recent reviews).

Recently, superradiant instabilities of ultralight particles around ECOs, have also been studied in Refs. \citet{guo2022near, Guo:2023mel}. In particular, a first effort was made
in Ref.~\cite{guo2022near} in investigating massive scalar perturbations with a
modified boundary condition, using a {purely analytic} approach in the
non-relativistic regime where $\alpha \equiv \mu M\ll 1$, where again $\mu$ and $M$ are the
masses of the bosonic field and the BH, respectively. The goal of this paper is to study this
subject in more detail and, for the first time, explicitly show how the boundary reflection influences the growth of massive scalar perturbations around ECOs. 

This paper is organized as follows. In Sec.~\ref{sec:eigenvalue problem} we
derive the equations of motion for massive scalar perturbations and specify the
boundary conditions. At the end of that section a discussion can be found regarding the difference in the boundary conditions set by \citet{guo2022near} and by us. In Sec.~\ref{sec:finding the growth rate} we solve the eigenvalue problem. We first
introduce our analytic method, from which we obtain our key result, the
correction factor $g_{\mathcal{K}}$ [see Eq.~(\ref{eq:gK}) below]. We also
calculate the eigenfrequencies using a semi-analytic method and a continued
fraction method, followed by a comparison between different methods. The
physical meaning of the correction factor $g_{\mathcal{K}}$ is investigated in
Sec.~\ref{section:physical interpretation} by making use of the energy-momentum
conservation. Sec.~\ref{sec:time evolution} is devoted to a discussion on the
time evolution of superradiant instabilities around ECOs, with a particular
focus on the influence of the boundary reflection. Sec.~\ref{sec:implications}
discusses the implications of the possible boundary reflection for the
constraints on the ultralight particle mass. Our summary and conclusion can be
found in Sec.~\ref{sec:summary}. In this work, We use the $(-, +, +, +)$
convention and set $G=c=\hbar=1$.

\section{Bosonic Cloud around an ECO}
\label{sec:eigenvalue problem}

We consider massive scalar perturbations in the following spacetime background: we assume that the geometry outside of the ECO is described by the Kerr metric,
with the line element in Boyer-Lindquist coordinates
\begin{equation}
\begin{aligned}
\dd s^2=&-\Big(1-\frac{2Mr}{\rho^2}\Big)\dd t^2+\frac{\rho^2}{\Delta}\dd
r^2-\frac{4Mr}{\rho^2}a \sin^2\theta \dd \phi \dd t\\ &+\rho^2\dd
\theta^2+\left[\big(r^2+a^2\big)\sin^2\theta+\frac{2Mr}{\rho^2}a^2\sin^4
\theta\right]\dd\phi^2,
\end{aligned}
\label{eq:Kerr metric}
\end{equation}
where $a=J/M$ is the spin angular momentum normalized by the mass of the ECO,
$\rho^2 \equiv r^2+a^2\cos^2\theta$, and $\Delta \equiv r^2-2Mr+a^2$. For a Kerr
BH, the event horizon and the Cauchy horizon are located at
$r_{+}=M+\sqrt{M^2-a^2}$ and $r_{-}=M-\sqrt{M^2-a^2}$ respectively; $r_+$ is
normally taken to be the inner boundary in the study of superradiant
instabilities, with a {purely ingoing} boundary condition \cite{detweiler1980klein,
dolan2007instability, baumann2019spectra}. However, for an ECO, we replace the
event horizon with a reflective surface located at $r_0=r_+(1+\epsilon)$, where
$\epsilon\ll 1$. This surface reflects a portion of the ingoing wave, 
parameterized by the reflectivity $\mathcal{K}$.

In a curved spacetime, a test scalar field with mass $\mu$ satisfies the Klein-Gordon equation
\begin{equation}
\big(\nabla^{\nu}\nabla_{\nu}-\mu^2 \big)\Psi=0.
\end{equation}
To study characteristic modes of the perturbation field, we separate variables
as follows
\begin{equation}
\Psi(t,r,\theta,\phi)=\ee^{-\ii\omega t}\ee^{\ii m\phi}R_{lm}(r)S_{lm}(\theta),
\label{eq:Psi}
\end{equation}
where $\omega\in\mathbb{C}$ is the complex eigenfrequency, and the integers $l \geq
0,\ m\in \left[-l,\ l\right]$ are the angular and magnetic quantum numbers,
respectively. Expanding the Klein-Gordon equation in a Kerr metric background, we
obtain the following equations of motion
\begin{widetext}
\begin{eqnarray}
\frac{\dd}{\dd r}\bigg(\Delta\frac{\dd R_{lm}}{\dd r}\bigg)+\Bigg[
\frac{\omega^2 \big(r^2+a^2 \big)^2-4Mam\omega r+m^2a^2}{\Delta} -\big(
\omega^2a^2+\mu^2 r^2+\Lambda_{lm} \big) \Bigg] R_{lm}(r) &=0 ,
\label{eq:eomR} \\
\frac{1}{\sin \theta}\frac{\dd}{\dd\theta}\bigg( \sin\theta\frac{\dd S_{lm}}{\dd
\theta} \bigg)+\Bigg[a^2 \big(\omega^2-\mu^2 \big)\cos^2\theta 
-\frac{m^2}{\sin^2\theta}+\Lambda_{lm} \Bigg] S_{lm}(\theta) &=0 ,
\label{eq:eomS}
\end{eqnarray}
\end{widetext}
where $\omega$ and $\Lambda_{lm}$ are the eigenvalues to be solved. The
eigenfunctions $S_{lm}$ of Eq.~(\ref{eq:eomS}) are a series of spin-weighted
spheroidal harmonics labelled by $l$ and $m$, with the eigenvalues
$\Lambda_{lm}=l(l+1)+\mathcal{O}\left[a^2(\mu^2-\omega^2)\right]$; see
Ref.~\cite{berti2006eigenvalues} for an analytical expansion of $\Lambda_{lm}$
in terms of $a\sqrt{\mu^2-\omega^2}$. Here we introduce a complex number
$l^{\prime}$ to denote $\Lambda_{lm}$ as
\begin{equation}
\Lambda_{lm}=l^{\prime}(l^{\prime}+1).
\end{equation}
When $\omega \approx \mu$ and $\alpha=\mu M\ll 1$ are satisfied, $l^{\prime}$ is
very close to the augular quantum number $l$. Therefore, in Eq.~(\ref{eq:eomR}), $\Lambda_{lm}$ can be treated as a known number and only the eigenfrequency $\omega$
needs to be found.

In order to find $\omega$ from Eq.~(\ref{eq:eomR}), we also need to impose
appropriate boundary conditions. When $r\to\infty$, we take the decaying
solution of $R_{lm}$ \cite{dolan2007instability}
\begin{equation}
\lim\limits_{r\to\infty} R_{lm}(r) \sim r^{-1+ \big( 2\omega^2-\mu^2
\big)M/\kappa} \ee^{-\kappa r},
\label{eq:outer boundary}
\end{equation}
where
\begin{equation}
\kappa =\sqrt{\mu^2-\omega^2},\quad\operatorname{Re} \kappa >0.
\label{eq:kappa}
\end{equation}

When investigating the behaviour of $R_{lm}$ near the inner boundary $r_0$, it
is useful to introduce the tortoise coordinate
\begin{equation}
\begin{aligned}
r^{\ast}&=\int\frac{r^2+a^2}{\Delta}\dd r\\ 
&=r+\frac{2Mr_+}{r_+-r_-}\ln|r-r_+|-\frac{2Mr_-}{r_+-r_-}\ln|r-r_-|,
\end{aligned}
\label{eq:tortoise coordinate}
\end{equation}
and define 
\begin{equation}
Y=(r^2+a^2)^{1/2}R_{lm}(r).
\end{equation}
The location of the boundary can be expressed in the tortoise coordinate,
$r_0^{\ast}=r^{\ast}(r_0)$.
Then the radial equation (\ref{eq:eomR}) can be rewritten in the standard form
of a wave equation ${\dd^2 Y}/{\dd r^{*2}}+VY=0$, where the effective potential
reads
\begin{equation}
\begin{aligned}
 V(r) =& -\frac{\Delta (2Mr^3+a^2r^2-4Ma^2r +a^4)}{(r^2+a^2)^4} \\ 
 & -\frac{\Delta(\mu^2r^2+a^2\omega^2-2ma\omega+\Lambda_{lm})}{(r^2+a^2)^2} +
 \Big(\omega-\frac{ma}{r^2+a^2}\Big)^2 \, .
\end{aligned}
\end{equation}

Since $r_0\approx r_+$, when $r\to r_0$ we have $\Delta\approx 0$, and therefore, in this limit,
the two independent solutions of $Y$ are $\ee^{\pm\ii(\omega-\omega_c)\Delta
r^{\ast}}$, where $\Delta r^{\ast}=r^{\ast}-r_0^{\ast}$, and
\begin{equation}
\omega_c=\frac{ma}{r_+^2+a^2}.
\label{eq:omegac}
\end{equation}
For an ECO, the wave function near $r_0$ should be a superposition of ingoing
and outgoing waves~\cite{Maggio:2019zyv}
\begin{equation}
\lim\limits_{\Delta r_{\ast}\to 0} Y\sim \ee^{-\ii(\omega-\omega_c)\Delta
r^{\ast}}+\mathcal{K}\ee^{\ii(\omega-\omega_c)\Delta r^{\ast}}
\label{eq:inner boundary}
\end{equation}
where $\mathcal{K}$ is the boundary reflectivity; $|\mathcal{K}|$ denotes the
proportion of the incident wave reflected at $r_0$, and $\arg(\mathcal{K})$ is
the phase shift.

Note that the inner boundary condition (\ref{eq:inner boundary}) in our
treatment differs from Eq.~(27) in Ref.~\cite{guo2022near}. In the latter, they
wrote $\ee^{-\ii(\omega-\omega_c)r^{\ast}} +
\mathcal{R}(\omega)\ee^{\ii(\omega-\omega_c)r^{\ast}}$ and the location
$r_0^{\ast}$ of the reflective surface is not explicitly specified. Their
definition is related to ours by
$\mathcal{R}(\omega)=\mathcal{K}\ee^{-2\ii(\omega-\omega_c)r_0^{\ast}}$. With
our definition, the physical meaning of the reflectivity $\mathcal{K}$ is
clearer. At the reflective boundary, we have $\Delta r^{\ast}=0$, and the
amplitudes of the ingoing and outgoing waves at the reflective boundary are $1$
and $\mathcal{K}$ respectively, up to a common constant. This means that
$|\mathcal{K}|$ is simply the reflected proportion, and $|\mathcal{K}|=1$
represents a ``perfect reflection'', up to a phase shift if $\mathcal{K}$ is
complex.

Finally, the radial equation (\ref{eq:eomR}), together with Eq.~(\ref{eq:outer
boundary}) and Eq.~(\ref{eq:inner boundary}), defines an eigenvalue problem.
This will be solved in the following where we will compute the complex eigenfrequencies
$\omega=\omega_R+\ii \omega_I$, where the real part $\omega_R$ denotes the energy
level of the bosonic cloud and the imaginary part $\omega_I$ represents
the growth rate of the superradiant instability.

\section{Growth rate of superradiant instability}
\label{sec:finding the growth rate}

\subsection{Analytic method}

In the ``non-relativistic'' regime $\alpha=\mu M\ll1$, we have $\omega\approx
\mu$ and can solve Eq.~(\ref{eq:eomR}) using matched asymptotic expansions. This
approach, found by \citet{detweiler1980klein}, has been used to study
superradiant instabilities around BHs \cite{furuhashi2004instability,
pani2012perturbations, baumann2019spectra, bao2022improved} and was recently
extended by \citet{guo2022near} to account for a boundary reflection. Here we
further develop it with an ECO boundary condition. This can be done by solving
the radial equation in the ``far'' and ``near'' regions respectively, and
matching the two solutions in the ``overlap'' region.

In the far region where $r\gg M$, to leading order in $\alpha$,
Eq.~(\ref{eq:eomR}) can be written as, 
\begin{equation}
\frac{\dd^2 \left(rR\right)}{\dd r^2}+\left[(\omega^2-\mu^2)+\frac{2 M
\mu^2}{r}-\frac{l\p(l\p+1)}{r^2} \right]\left(rR\right)=0,
\label{eq:fareq}
\end{equation}
where we have dropped subscripts ($l,m$) for notation simplicity. Following
\citet{detweiler1980klein}, we define
\begin{equation}
\nu=M\mu^2/\kappa,
\label{eq:nu}
\end{equation}
where $\kappa$ was defined in Eq.~(\ref{eq:kappa}). Then the solution to
Eq.~(\ref{eq:fareq}) with the decaying boundary condition at infinity reads
\begin{equation}
R_{\text{far}}(r)=(2\kappa r)^{l\p} \ee^{-\kappa r}U(l\p+1-\nu,2l\p+2;2\kappa r),
\label{eq:Rfar}
\end{equation}
where $U(a,b,c;x)$ is the Tricomi confluent hypergeometric function with repect
to $x$. If $l\p+1-\nu=-n$ is an integer, $U(a,b,c;x)$ reduces to a polynomial,
which, in quantum mechanics, corresponds to eigenstates of hydrogen atoms, with
$n=0, 1, 2\cdots$ being the radial quantum number. However, since the inner
boundary condition is different from that of hydrogen atoms, we must introduce a
small deviation $\delta\nu\in\mathbb{C}$, via
\begin{equation}
\nu = l\p+n+1+\delta\nu,
\label{eq:delta nu}
\end{equation}
and the eigenfrequency is
\begin{equation}
\omega=\mu\sqrt{1-\frac{\alpha^2}{\nu^2}}=\omega_R+\ii\omega_I.
\label{eq:omega}
\end{equation}
The latter equation is derived from the definition of $\nu$ in
Eq.~(\ref{eq:nu}) directly.

We now explore the solution of $R(r)$ in the near region $r\sim r_0$. Here we
introduce a dimensionless distance 
\begin{equation}
z \equiv \frac{r-r_+}{r_+-r_-},
\label{eq:z}
\end{equation}
and the location of the inner boundary is
\begin{equation}
z_0 \equiv \frac{r_0-r_+}{r_+-r_-}.
\label{eq:z0}
\end{equation}
Then, to leading order in $\alpha$, Eq.~(\ref{eq:eomR}) can be written as
\begin{equation}
z(z+1)\frac{\dd}{\dd z}\left[z(z+1)\frac{\dd R}{\dd z}\right]+V(z)R=0,
\label{eq:neareq}
\end{equation}
where\footnote{Our definition of $p$ differs from $P$ in
\citet{detweiler1980klein} by a minus sign.}
\begin{eqnarray}
V(z) &=& p^2-l\p(l\p+1)z(z+1), \\
p &=& \frac{2Mr_+(\omega-\omega_c)}{r_+-r_-}, \label{eq:p}
\end{eqnarray}
and $\omega_c$ is defined in Eq.~(\ref{eq:omegac}). The solution is
\begin{equation}
R_{\text{near}}(z)=\left(\frac{z}{z+1} \right)^{-\ii p}G(-l\p,l\p+1;1+2\ii p;z+1),
\label{eq:Rnear G}
\end{equation}
where $G(a,b;c;x)$ is any solution to the hypergeometric equation with respect to
$x$. There are two independent solutions~\cite{bateman1953higher}\footnote{In
Ref.~\cite{detweiler1980klein} the term corresponding to $2l\p+2$ is incorrectly
written as $2l+1$.},
\begin{equation}
\begin{aligned}
u_3&=(-z)^{l\p}\tensor[_2]{F}{_1}(-l\p,-l\p+2\ii p;-2l\p;-z^{-1}),\\
u_4&=(-z)^{-l\p-1}\tensor[_2]{F}{_1}(l\p+1,l\p+1+2\ii p;2l\p+2;-z^{-1}),
\label{eq:u34}
\end{aligned}
\end{equation}
where $\tensor[_2]{F}{_1}$ is the hypergeometric function. With these two
solutions, $R_{\text{near}}$ can be written as
\begin{equation}
R_{\text{near}}=\left(\frac{z}{1+z}\right)^{-\ii p}(b_3u_3+b_4 u_4).
\label{eq:Rnear u34}
\end{equation}
The ratio of the coefficients $b_3, b_4$ is determined by the inner boundary
condition~(\ref{eq:inner boundary}). Details are provided in 
Appendix~\ref{appendix: ratiob43}, and the result is
\begin{widetext}
    \begin{equation}
    \frac{b_4}{b_3}=-\frac{\Gamma (-2 l\p) \Gamma (l\p+1)}{\Gamma (-l\p) \Gamma (2 l\p+2)}
    \frac{\mathcal{K}z_0^{-2 i p}  \Gamma (2 i p+1) \Gamma (l\p-2 i p+1)-\Gamma (1-2 i p)
       \Gamma (l\p+2 i p+1)}{\mathcal{K}z_0^{-2 i p}  \Gamma (2 i p+1) \Gamma (-l\p-2 i
       p)-\Gamma (1-2 i p) \Gamma (2 i p-l\p)}.
    \label{eq:ratiob43}
    \end{equation}
\end{widetext}

Let us note that, to obtain Eq.~(\ref{eq:neareq}),
besides $\alpha\ll 1$, we have implicitly assumed two additional conditions by
neglecting terms at higher orders in $\alpha$,\footnote{These two conditions are
found by comparing the dominant term $p^2-\Lambda_{lm}z(z+1)$ with subdorminant
terms at $\mathcal{O}(\alpha^2)$ in the full expression of $V(z)$, which was
given in Ref.~\cite{bao2022improved}.} namely
\begin{align}
	\alpha^2(1-\omega_c/\omega) &\ll l(l+1)\sqrt{1-a^2/M^2} , \\
	z &\ll \min\left(l^2/\alpha^2,l/\alpha \right) .
\end{align}
 The former condition may not be satisfied even for $\alpha\ll 1$ if the spin is
 extreme $a/M \sim 1$, requiring the inclusion of the next-to-leading order
 correction for highly spinning BHs \cite{bao2022improved}. Here we do not include this correction for
 simplicity. The latter condition gives the regime of validity of the near
 region solution~(\ref{eq:Rnear u34}). 

The far region with $r\gg M$ and the near region with $z\ll
\min\left(l^2/\alpha^2,l/\alpha \right)$ have an overlap when $\alpha$ is small
enough, and thus the two solutions can be matched.

\begin{widetext}
    
First, one can expand the far region solution in the small-$r$ limit, keeping
only dominant terms:
\begin{equation}
    R_{\text{far}}(r)  =  (-1)^n
    \frac{\Gamma(2l\p+2+n)}{\Gamma(2l\p+2)}(2\kappa r)^{l\p} 
     +  (-1)^{n+1}\delta\nu \Gamma(2l\p+1) \Gamma(n+1) (2\kappa
    r)^{-l\p-1},
\label{eq:farexpansion}
\end{equation}
where $\kappa r\ll1$ and $|\delta\nu|\ll1$. Also, the large-$r$ limit of the
near region solution is
\begin{equation}
R_{\text{near}}(r) = b_3 \left(\frac{-r}{r_+-r_-}\right)^{l\p} + b_4
\left(\frac{-r}{r_+-r_-}\right)^{-l\p-1}.
\end{equation}
The two expansions should be linearly dependent, which determines $\delta\nu$
and $\omega$.

When doing the calculation, we make use of the fact that $l\p\approx l$ so only
$l$ appears in the final results. However, the factor
$\Gamma(-2l\p)/\Gamma(-l\p)$ needs to be treated with care. As in
Ref.~\cite{bao2022improved}, we take the limit $l\p\to l$, and it becomes
\begin{equation}
\lim_{l\p\to l}\frac{\Gamma(-2l\p)}{\Gamma(-l\p)} = \lim_{\epsilon\to
0}\frac{\Gamma(-2l-2\epsilon)}{\Gamma(-l-\epsilon)}=
\frac{(-1)^l}{2}\frac{\Gamma(l+1)}{\Gamma(2l+1)}.
\end{equation}
We also take into account that $|\delta\nu|\ll 1$, which allows us to obtain
$\omega$ with a Taylor expansion. Finally, we obtain
\begin{eqnarray}
M\omega_{R} &=& {\alpha\left[1-\frac{\alpha^2}{2(l+n+1)^2}\right]},
\label{eq:omega_R} \\
 M\omega_{I} &=&  g_{\mathcal{K}} \alpha^{4l+5} \left(\frac{ma}{2M}-\omega_R r_+\right)
 \frac{2^{4l+2}(2l+n+1)!}{(l+n+1)^{2l+4}n!}
 \left[\frac{l!}{(2l+1)!(2l)!}\right]^2    \prod_{j=1}^{l}\left[j^2
 \bigg(1-\frac{a^2}{M^2} \bigg)+\bigg(2r_+\omega_R-\frac{ma}{M} \bigg)^2\right],
\label{eq:omega_I}
\end{eqnarray}
where we have defined 
\begin{equation}
g_{\mathcal{K}} = \frac{1-\left|\mathcal{K}\right|^2}{1 +
\left|\mathcal{K}\right|^2 + 2\operatorname{Re}(A^2z_0^{-2\ii
p}\mathcal{K})/\left|A\right|^2},
\label{eq:gK}
\end{equation}
where
$A \equiv \prod_{j=1}^{l}(j-2\ii p)$ with $p$ defined in Eq.~(\ref{eq:p}).

\end{widetext}

Equation~(\ref{eq:gK}) is our key result. The most important feature of our
analytic result is that the growth rate $M\omega_I$ when including a (partially) reflective boundary
condition differs from that of the BH case only by the factor $g_{\mathcal{K}}$.
This factor does not alter the superradiance condition (\ref{eq:SR condition}),
namely that when $\omega_R<ma/2Mr_+$, the scalar field extracts energy and
angular momentum from the ECO, growing exponentially.

When $\mathcal{K}=0$ we have $g_{\mathcal{K}}=1$, and the growth rate recovers
the BH case, which was found by \citet{detweiler1980klein} (except for a $1/2$
factor\footnote{Our result (\ref{eq:omega_I}), when $g_{\mathcal{K}}=1$, turns
out to be the half of the the growth rate obtained in
Ref.~\cite{detweiler1980klein}. This could be explained by a missing $1/2$
factor that should have been on the right-hand side of Eq.~(23) in
Ref.~\cite{detweiler1980klein}, possibly stemming from an inappropriate
treatment of $\Gamma(-2l\p)/\Gamma(-l\p)$. This $1/2$ factor is also discussed
in Refs.~\cite{pani2012perturbations, bao2022improved}. As a comparison, our
result agrees with Eq.~(2.32) of Ref.~\cite{baumann2019spectra}.}) and other
studies (see e.g., Ref.~\cite{baumann2019spectra}). The factor $g_{\mathcal{K}}$
hence represents the correction introduced by the boundary reflection. The denominator of this factor can be written as
\begin{equation}
    1 + 
\left|\mathcal{K}\right|^2 + 2\left|\mathcal{K}\right|\cos\varphi,
\end{equation}
where $\varphi\equiv 2 \sum_{j=1}^{l}\arctan\left(-2p/j\right)-2p\ln z_0+\arg\mathcal{K}$. When $p$ changes, the denominator oscillates between $\big(1-\left|\mathcal{K}\right|\big)^2$ and $\big(1+\left|\mathcal{K}\right|\big)^2$. Therefore, we have

\begin{equation}
\frac{1-\left|\mathcal{K}\right|}{1+\left|\mathcal{K}\right|} \leq g_{\mathcal{K}} \leq
\frac{1+\left|\mathcal{K}\right|}{1-\left|\mathcal{K}\right|}.
\end{equation}
The value of $z_0$ will influence the (quasi-)period of the oscillation. For example, when $z_0$ is small enough so that $-2p\ln z_0$ is the dominant term in $\varphi$, the change in $p$ that accounts for a full oscillation cycle is $\sim \pi/\left|\ln z_0\right|$. In this paper we will illustrate results obtained with $z_0=10^{-5}$, and smaller values of $z_0$ will result in denser oscillatory patterns.

The physical meaning of $g_{\mathcal{K}}$ will be discussed in 
Sec.~\ref{section:physical interpretation}.

\subsection{Semi-analytic method}

The above method can yield an analytic result and clearly show how the boundary
reflection changes the growth rate. However, to get more accurate results,
the following semi-analytic method may be adopted, which was also used in Ref.~\cite{arvanitaki2011exploring} and similar to the method used in Ref.~\cite{rosa2010extremal}.
In the matching procedure presented above, only terms proportional to $r^{l\p}$
and $r^{-l\p-1}$ were considered. Therefore, a natural improvement to this scheme
is to compute Eq.~(\ref{eq:Rfar}) and Eq.~(\ref{eq:Rnear u34}) numerically and
match the two at a point $r_{\text{match}}$ in the overlapping region, via
\begin{equation}
\left(R_{\text{near}}\frac{\dd R_{\text{far}}}{\dd r}-R_{\text{far}}\frac{\dd
R_{\text{near}}}{\dd r}\right)\Bigg|_{r=r_{\text{match}}}=0.
\label{eq:Wronskian match}
\end{equation}
Since Eq.~(\ref{eq:Rfar}) and Eq.~(\ref{eq:Rnear u34}) are both approximate
solutions, one should find nonzero residuals after plugging them into the
original radial equation~(\ref{eq:eomR}). The point $r_{\text{match}}$ is chosen
such that relative residuals of the two solutions are equal or closest. This approach makes
use of the analytic solutions, $R_{\text{far}}$ and $R_{\text{near}}$, but
matches them numerically. Therefore, the method is semi-analytic.

\begin{figure}[!htbp]
\centering
\includegraphics[width=0.48\textwidth]{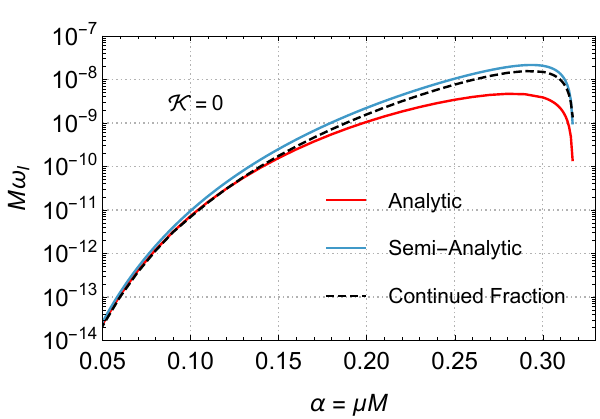}

\includegraphics[width=0.48\textwidth]{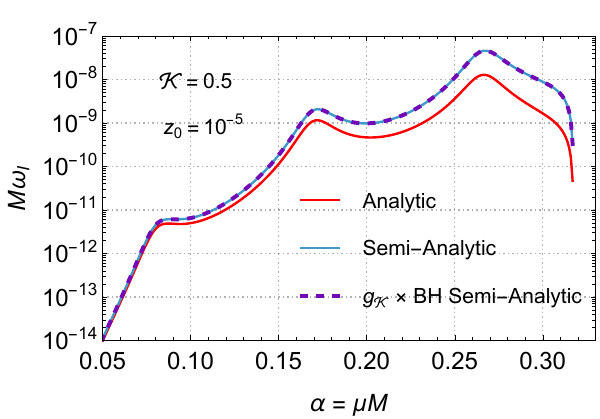}

\includegraphics[width=0.48\textwidth]{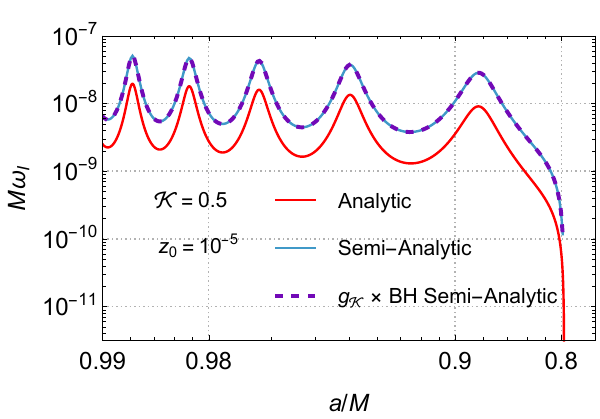}

\caption{Comparison of different methods in calculating the (dimensionless)
superrandiance growth rate $M\omega_I$. Growth rates $M\omega_I$ of the
fundamental mode ($l=m=1$ and $n=0$), calculated using the analytic method and
the semi-analytic method, are shown for BHs ({\it upper}) and ECOs ({\it middle}
and {\it bottom}), with $a/M=0.9$ for the first two panels and $\mu M=0.25$ for the last. For the BH case, we also present the result calculated using the continued
fraction method for comparison. For the ECO cases, additionally, we multiply the semi-analytic results for their corresponding BHs by $g_{\mathcal{K}}$, and plot the resulting growth rates with dashed purple curves.}
\label{fig:comparing methods}
\end{figure}

In Fig.~\ref{fig:comparing methods} we perform a comparison of the growth rates
calculated with different methods. We first consider the case of
$\mathcal{K}=0$, which corresponds to the {purely ingoing} boundary condition for a
BH. In this case, other than the analytic method and the semi-analytic method
explained above, the growth rate $\omega_{I}$ can also be calculated using the
continued fraction method \cite{dolan2007instability}, which serves as a
consistency check here. For our choice of parameters, the analytic results agree
very well with the continued fraction results in the regime $\alpha\ll 1$, but
the discrepancy quickly increases for a larger $\alpha$. However, the
semi-analytic method always yields a result close to that of the continued
fraction method for the full range of $\alpha$ considered, with a relative error
less than $50\%$. 

When $\mathcal{K}\neq 0$, the continued fraction method by
\citet{dolan2007instability} cannot be used, because the inner boundary
condition is no longer {purely ingoing}. In this case, in order to relate the semi-analytic results for ECOs to those for BHs, we also show a third
curve in dashed purple in each of the last two panels of Fig.~\ref{fig:comparing
methods}. This curve was obtained by multiplying the semi-analytic results for BHs by
$g_{\mathcal{K}}$. The striking point is, it perfectly agrees with the
semi-analytic results for ECOs, with a relative error less than $10^{-3}$.
Therefore, even though the correction factor $g_{\mathcal{K}}$ was obtained using the
fully analytical approach in the regime $\alpha\ll 1$, it turns out to also be applicable to
the more accurate semi-analytical results, which is not limited to that regime.

\begin{figure*}
\centering
\begin{minipage}[t]{0.48\textwidth}
\vspace{0pt}
\includegraphics[width=\textwidth]{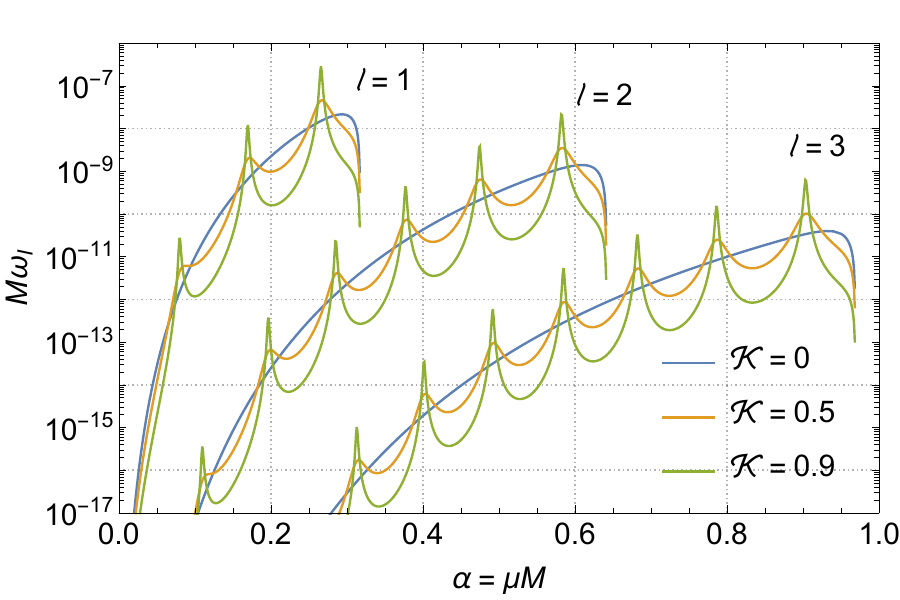}
\end{minipage}
\quad
\begin{minipage}[t]{0.48\textwidth}
\vspace{0pt}
\includegraphics[width=0.989\textwidth]{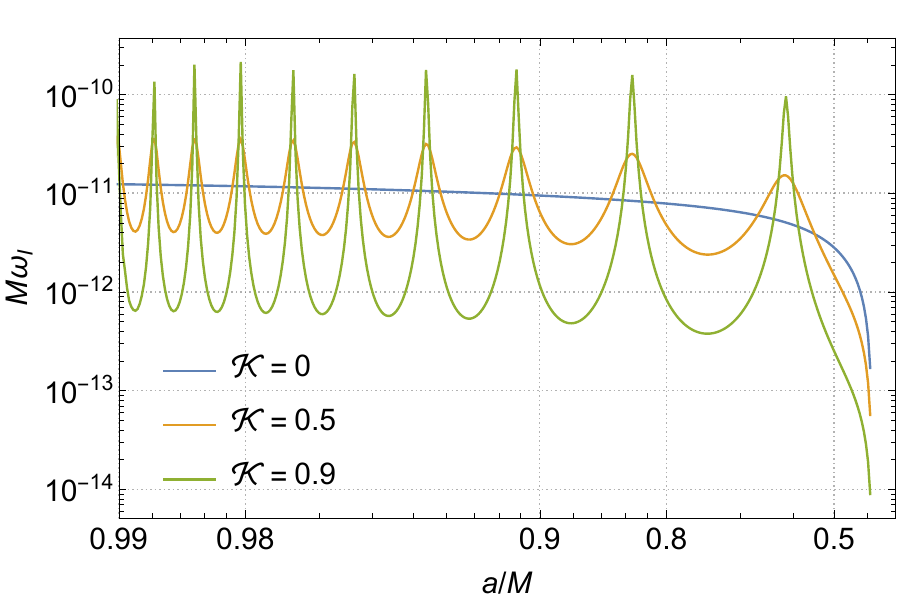}
\end{minipage}
\caption{Growth rate calculated using the semi-analytic method, shown as
functions of ({\it left}) the scalar mass parameter $\alpha$ and ({\it right}) the ECO's (dimensionless) spin $a/M$. The inner
boundary is located at $z_0=10^{-5}$. For the left panel, we take $a/M=0.9$ and
plot the modes $(l,m,n)=(1,1,0)$, $(2,2,0)$, and $(3,3,0)$. For the right, we
take $\mu M=0.1$ and only the mode $(l, m, n)=(1,1,0)$ is shown.}
\label{fig:growth rate}
\end{figure*}

In the last two panels of Fig.~\ref{fig:comparing methods}, the growth rate
exhibits oscillatory behaviors, introduced by the boundary reflection. In order
to see how the value of the reflectivity changes these  oscillating patterns, in Fig.~\ref{fig:growth
rate} we present the growth rate for different $\mathcal{K}$, calculated using
the semi-analytic method. It can be seen that for a larger $|\mathcal{K}|$, the
curve shows sharper peaks with deeper valleys in between. These new features
can affect the time evolution of superradiant instabilities and the ultralight
particle mass constraints. These will be investigated in Sec.~\ref{sec:time
evolution} and Sec.~\ref{sec:implications}; but before that, we examine the
physical origin of the correction factor $g_{\mathcal{K}}$ in the next section.

\section{Physical interpretation of $g_{\mathcal{K}}$}
\label{section:physical interpretation}

Here we try to understand Eq.~(\ref{eq:omega_I}) and the correction factor
$g_{\mathcal{K}}$ by analyzing the energy-momentum conservation at the ECO's
surface $r=r_0$. A similar analysis for BHs can be found in
Ref.~\cite{dolan2007instability}.

A complex scalar field has a Lagrangian density $\mathcal{L} =
\frac{1}{2}(\partial^{\rho} \Psi\partial_{\rho} \Psi^{\ast}+\mu^2
\Psi^{\ast}\Psi)$, and an energy-momentum tensor $T^{\mu\nu} =
\partial^{(\mu}\Psi\partial^{\nu)}\Psi^{\ast} - g^{\mu\nu}\mathcal{L}$.
Following \citet{dolan2007instability}, we use the ingoing-Kerr coordinates
$\tilde{x}^{\mu}=(\tilde{t}, r, \theta, \tilde{\phi})$, defined via
\begin{equation}
\tilde{t}=t+\alpha(r), \qquad \tilde{\phi}=\phi+\beta(r),
\end{equation}
where
\begin{equation}
\begin{aligned}
\alpha(r)&=\frac{2M}{r_+-r_-}\Big(r_+\ln|r-r_+|-r_-\ln|r-r_-|\Big),\\ 
\beta(r)&=\frac{a}{r_+-r_-}\ln\left|\frac{r-r_+}{r-r_-}\right|.
\end{aligned}
\end{equation}
Hereafter we add a tilde on top of quantities calculated in this coordinate
system. The contravariant metric tensor is
\begin{equation}
\tilde{g}^{\mu\nu}=\frac{1}{\rho^2}\left(
\begin{matrix}
-\rho^2-2Mr & 2Mr & 0 & 0 \\ 
2Mr & \Delta & 0 & a \\ 
0 & 0 & 1 & 0 \\ 
0 & a & 0 & 1/\sin^2\theta   
\end{matrix}\right) ,
\end{equation}
where $\rho$ and $\Delta$ are the same as in Eq.~(\ref{eq:Kerr metric}). Note
that our convention differs from that of \citet{dolan2007instability} by a minus
sign. In this case, the Klein-Gordon equation is separable using
\begin{equation}
\widetilde{\Psi}(\tilde{t}, r, \theta, \tilde{\phi})=\ee^{-\ii \omega
\tilde{t}}\ee^{\ii m \tilde{\phi}}S_{lm}(\theta)\widetilde{R}_{lm}(r).
\label{eq:Psi in ingoing coordinates}
\end{equation}

The spacetime has a Killing vector $\partial_{\tilde{t}}$, and
$\tensor{\widetilde{T}}{_{0}^{\mu}}$ is the conserved energy flux. We consider
the spacetime region $V$ that describes a time slice of the external space,
satisfying $-\Delta\tilde{t}/2<\tilde{t}<\Delta\tilde{t}/2$, $r>r_0$,
$0\leq\theta<\pi$ and $0\leq\tilde{\phi}<2\pi$. Then the conservation law,
$\nabla_{\mu}\tensor{\widetilde{T}}{_0^{\mu}}=0$, together with Gauss's
theorem, gives
\begin{equation}
\int_{\partial V}\tensor{\widetilde{T}}{_0^{\mu}} \tilde{n}_{\mu}
\sqrt{|\tilde{g}|} \dd^3 \tilde{S}=0,
\label{eq:gauss}
\end{equation}
where $\tilde{g} \equiv -\rho^4\sin^2\theta$ is the determinant of the covariant
metric $\tilde{g}_{\mu\nu}$, and $\tilde{n}_\nu$ is the normal one-form of
$\partial V$. Here $\tilde{n}_\nu=\pm \delta_{\nu}^0$ for the hypersurfaces
$\tilde{t}=\pm \Delta \tilde{t}/2$, and $\tilde{n}_\nu=\delta_{\nu}^1$ for the
hypersurface $r=r_0$. The hypersurface at spatial infinity $r \to \infty$ is not
included because the energy flux is zero there. When $\Delta\tilde{t}\to 0$,
Eq.~(\ref{eq:gauss}) yields the energy conservation equation
\begin{equation}
\frac{\partial}{\partial \tilde{t}}\int\limits_{\text{{3D}}}
-\tensor{\widetilde{T}}{_0^{0}}\rho^2\sin\theta \, \dd r \, \dd \theta \,\dd
\tilde{\phi} =  \int\limits_{\text{{2D}}}  -\tensor{\widetilde{T}}{_0^1}
\rho^2\sin\theta \, \dd\theta \, \dd\tilde{\phi},
\label{eq:energy conservation}
\end{equation}
where the ``3D'' integration is done in the external space ($r>r_0$), and the
``2D'' integration on the surface $r=r_0$, both at the fixed time $\tilde{t}=0$.

In order to obtain the asymptotic behavior of $\widetilde{R}_{lm}$, we compare
Eq.~(\ref{eq:Psi}) and Eq.~(\ref{eq:Psi in ingoing coordinates}), and find that
the radial function in the ingoing-Kerr coordinates $\widetilde{R}_{lm}$ and
that in the Boyer-Lindquist coordinates $R_{lm}$ are related by
\begin{equation}
\widetilde{R}_{lm}(r)=\ee^{\ii \omega \alpha(r)}\ee^{-\ii m \beta(r)}R_{lm}(r).
\end{equation}
Therefore, when $r\to r_0$, the radial function $\widetilde{R}_{lm}$ behaves as
\begin{equation}
\widetilde{R}_{lm}\sim z^{\ii p}R_{lm}\sim z^{\ii p}\cdot
C_{lm}\left[(z/z_0)^{-\ii p}+\mathcal{K}(z/z_0)^{\ii p}\right],
\label{eq:asymptotic tildeR}
\end{equation}
where $C_{lm}$ is a constant. Since in our calculation, from Eq.~(\ref{eq:Rfar}) to the subsequent matching procudure, the absolute magnitude of the field is not specified, we call $C_{lm}$ the ``relative amplitude'' of the field at the inner boundary.\footnote{Strictly speaking, the value of the radial function $R_{lm}$ at $r=r_0$ is $C_{lm}(1+\mathcal{K})$.} In most
cases, $|C_{lm}|^2\ll 1$. Using the transformed wave function $\widetilde{\Psi}$ (\ref{eq:Psi in ingoing coordinates}) and the asymptotic behavior of $\widetilde{R}_{lm}$ (\ref{eq:asymptotic tildeR}), direct calculation yields 
\begin{equation}
-\tensor{\widetilde{T}}{_0^1} = \frac{\omega_R
(ma-2Mr_+\omega_R)}{\rho_0^2}\left|C_{lm}\right|^2
\Big(1-\left|\mathcal{K}\right|^2 \Big)\left|S(\theta)\right|^2,
\label{eq:T01}
\end{equation}
which is the net energy flux going outwards at $r_0$, defined in the ingoing-Kerr
coordinates. Calculating the 2D integral in Eq.~(\ref{eq:energy conservation}),
we obtain
\begin{equation}
\begin{aligned}
2 \omega_I = \omega_R \Big(1-\left|\mathcal{K}\right|^2 \Big)  \frac{\big(
ma-2Mr_+\omega_R \big) \left|C_{lm}\right|^2}{{\int\limits_{\text{{3D}}}
-\tensor{\widetilde{T}}{_0^{0}} \, \rho^2\sin\theta \, \dd r \, \dd \theta \,
\dd \tilde{\phi}}} ,
\label{eq:energy-momentum and growth rate}
\end{aligned}
\end{equation}
where the $2\omega_I$ term arises because $\tensor{\widetilde{T}}{_0^0}\propto
\ee^{2\omega_I \tilde{t}}$. The integral in the denominator represents the total
energy outside $r_0$. As long as $\left|C_{lm}\right|^2\ll1$, the integral
mainly depends on the far region solution $R_{\text{far}}(r)$ and therefore can
be approximately evaluated using a hydrogenic wave function in a Newtonian
potential. As a result, this integral has a very weak dependence on $a/M,
z_0$, and $\mathcal{K}$, and mainly depends on $l$ and $n$.

\begin{figure*}[!htbp]
\centering
\begin{minipage}[t]{0.48\textwidth}
\vspace{0pt}
\includegraphics[width=0.95\textwidth]{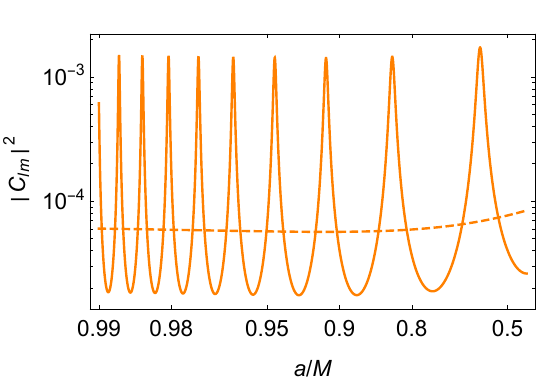}
\end{minipage}
\quad
\begin{minipage}[t]{0.48\textwidth}
\vspace{0pt}
\includegraphics[width=0.93\textwidth]{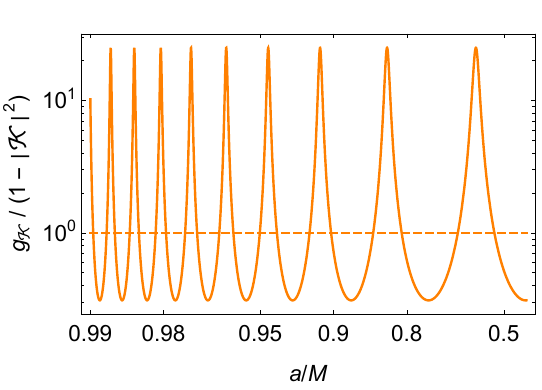}
\end{minipage}
\caption{A comparison between $\left|C_{lm}\right|^2$ and $g_{\mathcal{K}}/(1-\left|\mathcal{K}\right|^2)$. Solid lines are for an ECO with $\mathcal{K}=0.8$ and $z_0=10^{-5}$, while dashed lines are their counterparts
for a BH ($\mathcal{K}=0$). We take $\mu M=0.1$ and consider the fundamental mode $l=m=1$ and
$n=0$. The relative amplitude $C_{lm}$ is calculated using the analytic method.}
\label{fig:Clm}
\end{figure*}

Equations~(\ref{eq:T01}) and (\ref{eq:energy-momentum and growth rate}) provide a way to understand
the physical meaning of $g_{\mathcal{K}}$ in Eq.~(\ref{eq:gK}).  First, let us
consider the overall suppression factor $\big(1-|\mathcal{K}|^2\big)$. This
factor also appears in Eq.~(\ref{eq:T01}) and its
interpretation is straightforward: the energy flows carried by the ingoing
and outgoing waves go on opposite directions and thus the net flux is reduced when there is a (partially) reflecting surface. In particular, when $\left|\mathcal{K}\right|=1$, the two achieve a balance, and thus there is no net energy
flux across $r_0$, leading to a zero growth/decay rate.\footnote{Since the
denominator in Eq.~(\ref{eq:gK}) ranges from $\big(1-|\mathcal{K}|\big)^2$ to
$\big(1+|\mathcal{K}|\big)^2$, one may wonder what if $|\mathcal{K}|=1$ and both
the numerator and denominator are equal to zero. In this case, $g_{\mathcal{K}} \to
\infty$. However, once the spin $a$ decreases due to extraction of the angular
momentum, the denominator would be nonzero and  $g_{\mathcal{K}}$ stays at 0
thereafter.}

The denominator of $g_{\mathcal{K}}$, which represents the oscillatory behavior of $\omega_I$, is tightly related to the relative amplitude $C_{lm}$. To demonstrate this, in the left panel of Fig.~\ref{fig:Clm}, we plot $\left|C_{lm}\right|^2$ for $\mathcal{K}=0.8$ (solid) and $\mathcal{K}=0$ (dashed). For classical BHs, $|C_{lm}|^2$ changes very slowly with $a$. However, in the presence of a boundary reflection, the change becomes rapid. The oscillatory behaviour of $\left|C_{lm}\right|^2$ directly leads to the oscillatory behaviour of $\omega_I$ via Eq.~(\ref{eq:energy-momentum and growth rate}), manifested as the oscillating denominator of $g_\mathcal{K}$, as shown in the right panel of Fig.~\ref{fig:Clm}. The physical link between the relative amplitude and the growth rate is also straightforward: with a larger $|C_{lm}|^2$, the scalar field extracts a larger energy flux (\ref{eq:T01}) and thus grows faster.

To sum up, in the analytical expression of $g_{\mathcal{K}}$ (\ref{eq:gK}), the
factor $\big(1-|\mathcal{K}|\big)^2$ can be understood as the counteraction of
outgoing and ingoing energy flows, and the oscillatory behavior of the
denominator comes from the change in the scalar field's (relative) density $|C_{lm}|^2$ at
$r=r_0$, which is proportional to the amount of energy extracted there.

\section{Time evolution of superradiant instability}
\label{sec:time evolution}

In this section, we consider the time evolution of superradiant instabilities
and investigate how it is influenced by the boundary reflection.

Let us start by reviewing the case of a Kerr BH. If initially there is a nonzero
scalar field around a Kerr BH (for example, arising from quantum fluctuations),
as long as the superradiance condition (\ref{eq:SR condition}) is satisfied, the
field will extract energy and angular momentum from the BH. In this case, more
and more scalar particles are produced, the field grows exponentially, and a
bosonic cloud around the BH is formed. For scalar fields, the time evolution of superradiant
instabilities around Kerr BHs has been investigated using an adiabatic approximation in Ref.~\cite{brito2015black}.
For the case of an ECO, the superradiance condition is not changed. Similar to
its BH counterpart, a scalar field could grow and form a bosonic cloud when the
superradiance condition is satisfied. However, since the growth rate is changed
by the factor $g_{\mathcal{K}}$, one will anticipate some new features in the
time evolution. 

To begin with, we present the equations governing the adiabatic evolution of the instability. These equations
are essentially the same as in the case of a Kerr BH. First, since the wave
function $\Psi$ grows as $\sim \ee^{\omega_I t}$, the number of particles in the
bosonic cloud grows as $\sim \ee^{2\omega_I t}$. Therefore, the superradiant
energy extraction rate is
\begin{equation}
\dot{E}_\text{SR}=2\omega_I M_\text{cl},
\end{equation}
where $M_\text{cl}$ is the mass of the bosonic could. Since we are mostly interested in the case where gas accretion is much slower than the evolution of
superradiant instabilities, we will not take possible accretion processes onto the central ECO into
account.\footnote{See Ref.~\cite{brito2015black} where gas accretion is included
in the evolution equations for BHs.} Then the mass $M$ and angular momentum $J$
of the ECO change according to
\begin{align}
\dot{M} &=-\dot{E}_\text{SR},
\label{eq:ECO dM} \\
\dot{J} &=-\frac{m}{\omega_R}\dot{E}_\text{SR}.
\label{eq:ECO dJ}
\end{align}
On the other hand, the mass of the bosonic cloud changes as
\begin{equation}
\dot{M}_{\text{cl}}=\dot{E}_\text{SR}-\dot{E}_{\text{GW}},
\end{equation}
where $\dot{E}_{\text{GW}}$ is the energy flux carried away by GWs emitted by the cloud. In our study
below, we only consider the fundamental mode, $l=m=1$ and $n=0$, for which we
can adopt the GW energy flux obtained in Ref.~\cite{brito2015black},
\begin{equation}
\dot{E}_{\text{GW}} = \frac{484 +
9\pi^2}{23040}\left(\frac{M_{\text{cl}}^2}{M^2}\right)(M\mu)^{14}.
\end{equation}
This equation, obtained for the case of BHs, is  supposed to be a good
approximation still for the case of ECOs. The reason is that the GW emission
mostly comes from the far region where $|\Psi|^2$ is large, and the far region
wavefunctions in both the BH and the ECO cases are nearly the same.

\begin{table}[!htbp]
\centering
\caption{Definition of parameters for the time evolution of
superradiant instabilities.\label{table: parameters}}
\renewcommand\arraystretch{1.3}
\begin{tabular}{cc}
\hline\hline
Parameter & Definition  \\ 
\hline
$M_0$ & Initial mass of the ECO  \\ 
$M_{\text{cl},0}$ & Initial mass of the bosonic cloud \\ 
$J_0$ & Initial spin of the ECO \\ 
$\mu$ & Mass of the ultralight scalar particle \\ 
$z_0$ & Location of the reflective boundary in Eq.~(\ref{eq:z0})\\ 
$\mathcal{K}$ & Reflectivity of the boundary surface \\  \hline
\end{tabular}
\end{table}

\begin{figure}[!htbp]
    \centering
    \includegraphics[width=0.48\textwidth]{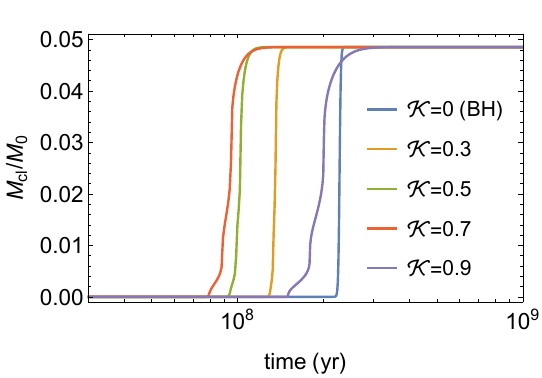}
    \includegraphics[width=0.46\textwidth]{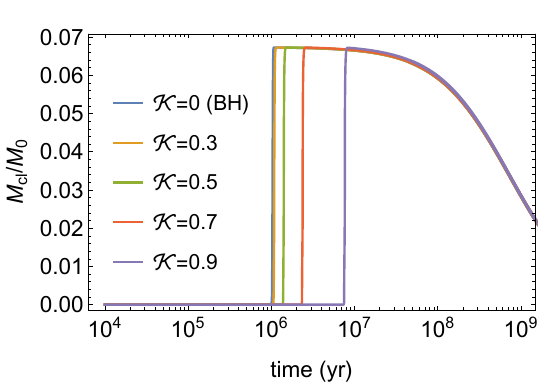}
    \caption{Time evolution of the mass of the bosonic cloud. We take $M_0=10^7
    \, M_{\odot},\ J_0/M_0^2=0.9,\ z_0=10^{-5}$, and only consider the
    fundamental mode with $l=m=1$ and $n=0$. We have used $\mu=\SI{1e-18}{eV}$
    and $\SI{2e-18}{eV}$ for the upper and lower panels respectively,
    corresponding to $\mu M=0.075$ and $\mu M=0.15$.}
    \label{fig:evolution}
\end{figure}

Using the parameters listed in Table \ref{table: parameters}, we can calculate
the time evolution of superradiant instabilities. Here the initial mass of the bosonic cloud
is taken to be the mass of a particle, $M_{\text{cl},0}=\mu$. The results are plotted in Fig.~\ref{fig:evolution} for different values of reflectivity and scalar particle mass. We found that the evolution can be roughly divided into three stages. 
\begin{description}
\item[Steady growth of the scalar field] In the very beginning, the mass of the
cloud is so small (in fact for $\mu=\SI{1e-18}{eV}$ and $M_0=10^{7}\ M_{\odot}$,
we have $M_{\text{cl},0}/M_0\sim 10^{-91}$) that the superradiant extraction is
negligible. Therefore, the mass and spin of the ECO are essentially unchanged
for a long period, and consequently, the growth rate $\omega_I$ stays steady.
\item[A spin-down phase of the ECO] After about $200$ e-folds, the cloud has
acquired a non-negligible mass $\sim 10^{-5} \, M_0$, and the evolution starts
to be discernible in the figure. The cloud quickly extracts energy and angular
momentum from the ECO, until it reaches the maximal mass $\sim 0.1 M_0$. This stage lasts for about $10$ e-folds,
during which the spin of the ECO drops quickly. Therefore, as implied in the right panel of
Fig.~\ref{fig:growth rate}, the growth rate $\omega_I$ shows an oscillatory
pattern with time. As an obvious example, the growth for the $\mathcal{K}=0.9$
case in the upper panel of Fig.~\ref{fig:evolution} is uneven during this stage.
\item[GW dissipation] After the spin of the ECO drops to the superradiant
critical value, GW emission overtakes the superradiant extraction, and therefore
the cloud starts to dissipate gradually. GWs emitted by the cloud are nearly
monochromatic with an angular frequency $\sim 2\omega_R$ and a slowly decreasing
amplitude \cite{arvanitaki2015discovering,Brito:2017zvb,siemonsen2023superrad}.
\end{description}

In Fig.~\ref{fig:evolution}, we can see that for different $\mathcal{K}$, the
time the cloud takes to accumulate to its maximal mass can vary. This timescale
is mainly determined by the first stage, during which the growth rate $\omega_I$
is essentially a constant. The boundary reflection changes this timescale via
the correction factor $g_{\mathcal{K}}$ in $\omega_I$. Since $g_{\mathcal{K}}$
can be either larger or less than $1$,  this timescale,  compared to the BH
case, can be either shortened or prolonged by the boundary reflection, as seen
in the upper (shortened) and lower (prolonged) panels of
Fig.~\ref{fig:evolution} respectively. The change in the growth timescale
implies that the boundary reflection may affect constraints on the ultralight
particle mass, which is the topic of the next section.

\section{Astrophysical constraints on the mass of ultralight bosons}
\label{sec:implications}

In this section, we first review how superradiant instabilities can be used to
constrain the mass of ultralight bosons, and then discuss the implications of
the reflective boundary condition if the assumed BH is in fact an ECO.

\subsection{Ultralight particle mass constraints from BHs}

Constraints on the mass of ultralight bosons have been imposed considering BH
superradiance, with measurements of BHs' spin and mass
\cite{arvanitaki2011exploring, arvanitaki2015discovering, arvanitaki2017black,
cardoso2018constraining}.  The basic idea is that, if there exists an ultralight
particle with mass $\mu$, BHs with high enough spins should suffer superradiant
instabilities and spin down. This results in an exclusion region on the
$J\text{--}M$ plane (Regge plane), where $J$ and $M$ are respectively the angular momentum and
mass of BHs. The location of this region is related to the mass of the particle
$\mu$. Therefore, values of $\mu$ which create exclusion regions that are
incompatible with existing BH $J\text{--}M$ measurements should not be allowed. 

To estimate the exclusion regions, one approach is to compare the characteristic
timescale of the bosonic cloud evolution, $\tau_{\text{cloud}}$, with the
characteristic timescale associated to the BH's astrophysical processes,
$\tau_{\text{astro}}$. For example, for the case of binary BHs,
\citet{arvanitaki2017black} compared the superradiance saturation timescale,
namely the time the cloud takes to accumulate to its maximal mass, with the
binary merger timescale. For the case of X-ray binaries,
\citet{cardoso2018constraining} compared the instability timescale,
$1/\omega_I$, with the durations over which two sources show stable spin values.
The typical astrophysical timescale is also often chosen to be the accretion
timescale of the BH \cite{arvanitaki2011exploring, pani2012perturbations,
PhysRevLett.109.131102, PhysRevD.96.035019}. If $\tau_{\text{cloud}} \ll
\tau_{\text{astro}}$, the cloud could extract energy and angular momentum
effectively within an astrophysical timescale, substantially spinning down the
BH.  Another approach to find the exclusion regions is the Monte Carlo method
\cite{brito2015black}. Starting with a particular $(J,M)$ combination, one can
calculate the evolution of the system and extract the final state of the BH at
some specified time $t_{F}$. Doing this for a sample of randomly chosen initial
states, $(J_i,M_i)$, and plotting the final states, $(J_f,M_f)$, on the Regge plane, one can find that a particular region is hardly populated;
see Fig.~3 in Ref.~\cite{brito2015black} for example.

\subsection{Extension to the ECO case and the role of boundary reflection}

Previous mass constraints of ultralight particles were obtained based on the
assumption that the compact objects are BHs, with an event horizon as the inner
boundary. However, it is worth studying how the change in the boundary condition
affects the mass constraints. Here we make a brief discussion on how the ECO
boundary condition alters the $J\text{--}M$ exclusion regions, as well as its
implications for ultralight particle mass constraints.

For simplicity, we adopt the first approach, i.e. comparing timescales, to draw
exclusion regions on the Regge plane. Here the astrophysical timescale
is taken to be the accretion timescale, $\tau_{\text{Acc}}$, of the ECO. We
assume that the ECO is accreting at a rate $f_{\text{Edd}}\dot{M}_{\text{Edd}}$,
where $\dot{M}_{\text{Edd}}$ is the Eddington accretion rate, which is related
to the Eddington luminosity $L_{\text{Edd}}$ through the radiative efficiency
$\eta$, via $ \epsilon \dot{M}_{\text{Edd}} c^2=L_{\text{Edd}}=1.26\times
10^{31} \big({M}/{M_{\odot}}\big) \, {\rm J\,s}^{-1}$.  The factor $\epsilon
\equiv \eta/(1-\eta)$ arises because if a fraction $\eta$ of the infalling mass
is converted to radiation, the accreted fraction reduces to $1-\eta$. Here we
define the accretion timescale
\begin{equation}
\tau_{\text{Acc}}\equiv
\frac{M}{f_{\text{Edd}}\dot{M}_{\text{Edd}}}=\frac{4.5\times 10^{7}}
{f_{\text{Edd}}}  \frac{\epsilon}{0.1}\ \si{yr}.
\end{equation}
where we shall typically take $\epsilon=0.1$ \cite{1964ApJ...140..796S,Shankar_2009}.

If we consider a supermassive ECO with mass
$M\sim 10^{6}\,M_{\odot}$, and take the initial mass of the cloud to be the mass
of a single scalar particle $\mu\sim 10^{-18}\ \si{eV}$, it takes the cloud
about $\ln(0.1M/\mu)\sim 205$ e-folds to grow to $0.1 \, M$. Therefore, we
define the fast-superradiance regime satisfying
\begin{equation}
205\ln\left(\frac{M}{10^6 M_{\odot}}\frac{10^{-18}\ \si{eV}}{\mu} \right)\
\tau_{\text{SR}}<\tau_{\text{Acc}},
\label{eq:fast SR}
\end{equation}
where the superradiance e-fold timescale is $\tau_{\text{SR}}\equiv
1/(2\omega_{I})$ since the cloud grows as $M_{\text{cl}}\propto \ee^{2\omega_I
t}$. In this regime, the cloud will extract energy and angular momentum
effectively within the accretion timescale. Therefore, it gives an exclusion
region on the Regge plane. BHs ($\mathcal{K}=0$) or ECOs
($\mathcal{K}\neq 0$) inside this region should spin down effectively and leave
this region within $\tau_{\text{Acc}}$.

We calculated the growth rate $\omega_{I}$ using the analytic method. Exclusion
regions defined in Eq.~(\ref{eq:fast SR}) are plotted in Fig.~\ref{fig:fast SR
regime} for different values of the boundary reflectivity $\mathcal{K}$. It is
clear that the boundary reflection alters the shape of this region and introduces
small spiky features, which are more pronounced as $|\mathcal{K}|$ increases.
When $\mathcal{K}$ approaches extremity $|\mathcal{K}|\to 1$, the spikes are
sharper, while the bulk part---the part under the spikes, as illustrated in the figure for $\mathcal{K}=0.99$---shrinks inwards. 

\begin{figure}[!htbp]
\centering
\includegraphics[width=0.48\textwidth]{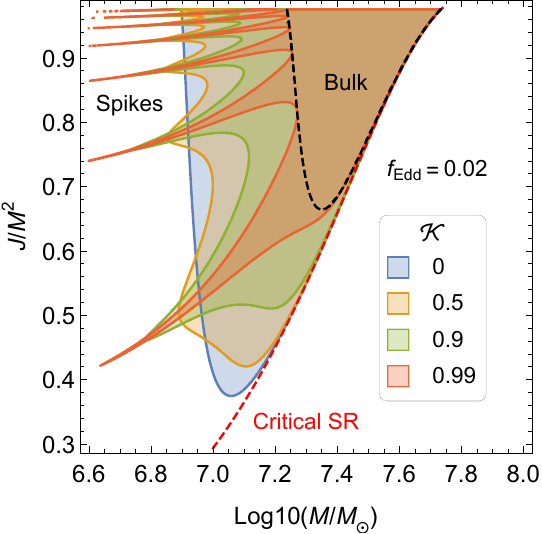}
\caption{Exclusion regions, as defined in Eq.~(\ref{eq:fast SR}),
for different $\mathcal{K}$. The red dashed line denotes where the superradiance
condition (\ref{eq:SR condition}) is saturated.  The vertical axis extends up to
$J/M^2=0.975$, above which the small spikes are much more crowded.  We have
taken $z_0=10^{-5}$ and $\mu=10^{-18}\ \si{eV}$.}
\label{fig:fast SR regime}
\end{figure}

Our results may have some implications for the usual method for constraining the
scalar mass. When the reflectivity is not too large, the alteration to the
exclusion region is insignificant. For example, in the $\mathcal{K}=0.5$ case of
Fig.~\ref{fig:fast SR regime}, compared with the $\mathcal{K}=0$ case, for every
value of $J/M^2$, the change in the  value of $M$ on the boundary line is within
$\SI{0.1}{dex}$, which can be comparable to, say, the current measurement
errors. However, for $\mathcal{K}=0.99$, the spikes are distinct, and the
shrinkage of the bulk region reaches $\sim\SI{0.4}{dex}$. Therefore, we expect
that a mildly reflective boundary, roughly $|\mathcal{K}|\lesssim0.5$, may not
substantially influence the mass constraints of ultralight scalar particles, but
an extreme value of reflectivity, say, $|\mathcal{K}|\gtrsim0.9$, could
introduce distinct spiky structures to the exclusion region, with a considerable
inward shrinkage of its bulk part.

\section{Conclusion}
\label{sec:summary}

Exotic compact objects (ECOs) have been conceived as alternatives to BHs. ECOs do not
possess an event horizon, and the inner boundary condition for scalar
perturbations is different from that of BHs. In this paper, we computed the
growth rate of superradiant instabilities assuming a modified inner boundary
condition, parameterized by the location of a reflective surface, $z_0$, and
its reflectivity, $\mathcal{K}$. We solved the eigenvalue problem analytically,
using matched asymptotic expansions, and found the analytic expression of the
growth rate $\omega_I$. Our key result is that the growth rate of superradiant
instabilities around an ECO can be related to the value in the BH case simply by
a factor $g_{\mathcal{K}}$, whose explicit expression is given in
Eq.~(\ref{eq:gK}). For a better accuracy, we also calculated the growth rate
using a semi-analytic method. We found that the semi-analytic results in the
ECO and BH cases can also be related by the same factor $g_{\mathcal{K}}$,
despite the fact that this factor was obtained using a {purely analytic} treatment.
Therefore, the factor $g_{\mathcal{K}}$ must have a clear physical meaning, which was investigated and we showed that it can be related to the energy flux at the inner boundary.

Using an adiabatic approach, we also studied how the superradiant instability of such ECOs would evolve. We found that, starting from a single particle, the evolution can be divided into three stages, namely (i)
steady growth of the scalar field, (ii) a spin-down phase of the ECO, and (iii) GW
dissipation. The time it takes for the cloud to reach its maximal mass mainly
depends on the duration of the first stage, and can be either shortened or
prolonged by the boundary reflection.

Finally, we discussed the implications for astrophysical constraints on
ultralight scalar fields. By comparing the timescales of the cloud
evolution and gas accretion, we found the exclusion regions on the ECOs'
Regge plane. Boundary reflection introduces spiky structures to the
exclusion region, and the effect is more pronounced for larger reflectivities.
As long as the reflectivity is not too large, say $|\mathcal{K}|\lesssim 0.5$,
the alteration to the exclusion region may not substantially influence the mass
constraints of ultralight scalars, but the effects of boundary reflection could
be significant for large reflectivity, e.g., $|\mathcal{K}|\gtrsim0.9$. 

{At the end of this paper, we make a short comment on the ECO model we adopted. Our work is based on the model in which one truncates the Kerr spacetime at a radius $r_0$ and puts a spherical reflective boundary there with an isotropic reflectivity $\mathcal{K}$. Although widely used in literature (as mentioned in the Introduction and references therein), this model is only a simplified one. More realistic models may consider deviation of the boundary shape from a sphere
and also anisotropic reflectivity, which is out of the scope of this work and deserves future study.}

\begin{acknowledgments}
This work was supported by the National Natural Science Foundation of China
(11991053, 12247128, 11975027), the National SKA Program of China
(2020SKA0120300), the Max Planck Partner Group Program funded by the Max Planck
Society, and the High-Performance Computing Platform of Peking University. L.Z.\
is supported by the Hui-Chun Chin and Tsung-Dao Lee Chinese Undergraduate
Research Endowment (Chun-Tsung Endowment) at Peking University. R.B. acknowledges financial support provided by FCT – Fundação para a Ciência e a Tecnologia, I.P., under the Scientific Employment Stimulus -- Individual Call -- 2020.00470.CEECIND and under project No. 2022.01324.PTDC.
\end{acknowledgments}

\appendix

\section{Determining the ratio $b_4/b_3$}
\label{appendix: ratiob43}

When $z\to 0$, we have
\begin{align}
\left(\frac{z}{1+z}\right)^{-\ii p}u_3 &\to f_3^{-}z^{-\ii p}+f_3^{+}z^{\ii p},\\ 
\left(\frac{z}{1+z}\right)^{-\ii p}u_4 &\to f_4^{-}z^{-\ii p}+f_4^{+}z^{\ii p},
\end{align}
where
\begin{align}
f_3^{-} &=\frac{(-1)^{l\p} \Gamma (-2 l\p) \Gamma (2 i p)}{\Gamma (-l\p) \Gamma (2 i p-l\p)},\\
f_3^{+} &=\frac{(-1)^{l\p} \Gamma (-2 l\p) \Gamma (-2 i p)}{\Gamma (-l\p) \Gamma (-l\p-2 i p)},\\ 
f_4^{-} &=\frac{(-1)^{1-l\p} \Gamma (2 l\p+2) \Gamma (2 i p)}{\Gamma (l\p+1) \Gamma (l\p+2 i p+1)},\\ 
f_4^{+} &=\frac{(-1)^{1-l\p} \Gamma (2 l\p+2) \Gamma (-2 i p)}{\Gamma (l\p+1) \Gamma (l\p-2 i p+1)}.
\end{align}

The inner boundary condition (\ref{eq:inner boundary}) is equivalent to
\begin{equation}
\lim\limits_{z\to z_0} R_{\text{near}}\sim (z/z_0)^{-\ii
p}+\mathcal{K}(z/z_0)^{\ii p}.
\label{eq:inner boundary z}
\end{equation}
Considering the boundary condition and the asymptotic behaviours of $u_3, u_4$,
we can pin down the ratio $b_4/b_3$ via
\begin{equation}
\frac{b_3 f_3^{+}+b_4 f_4^{+}}{b_3 f_3^{-}+b_4 f_4^{-}}=\mathcal{K}z_0^{-2\ii p},
\end{equation}
and the result is presented in Eq.~(\ref{eq:ratiob43}) in the main text.



\bibliography{refs}

\end{document}